\documentclass[a4paper,fleqn,usenatbib]{mnras}

\usepackage{graphicx}	
\usepackage{amsmath}	
\usepackage{amssymb}	

\newcommand{\g}{\gamma}

\title[Testing one-zone synchrotron-self-Compton models]{Testing one-zone synchrotron-self-Compton models with spectral energy distributions of Mrk 421}
\author[Zhu et al.]
{Qianqian Zhu$^{1}$,
Dahai Yan$^{2,3}$\thanks{E--mail: yandahai@ihep.ac.cn},
Pengfei Zhang$^{4}$,
Qian-Qing Yin$^{3,5}$,
Li Zhang$^{1}$\thanks{E--mail: lizhang@ynu.edu.cn},
\newauthor
Shuang-Nan Zhang$^3$\thanks{E--mail: zhangsn@ihep.ac.cn}\\
$^1$Department of Astronomy, Key Laboratory of Astroparticle Physics of Yunnan Province, Yunnan University, Kunming 650091, China\\
$^2$Yunnan Observatories, Chinese Academy of Sciences, Kunming 650011, China\\
$^3$Key Laboratory of Particle Astrophysics, Institute of High Energy Physics,
Chinese Academy of Sciences, Beijing 100049, China\\
$^4$Key Laboratory of Dark Matter and Space Astronomy, Purple Mountain Observatory,
Chinese Academy of Sciences, Nanjing 210008, China\\
$^5$University of Chinese Academy of Sciences, Beijing 100049, China}

\date{Accepted XXX. Received YYY; in original form ZZZ}

\pubyear{2016}

\begin{document}
\label{firstpage}
\pagerange{\pageref{firstpage}--\pageref{lastpage}}
\maketitle

\begin{abstract}
We test one-zone synchrotron self-Compton (SSC) models with high-quality multiwavelength spectral energy distribution (SED) data of Mrk 421.
We use Markov chain Monte Carlo (MCMC) technique to fit twelve day-scale SEDs of Mrk 421 with one-zone SSC models.
Three types of electron energy distribution (EED), a log-parabola (LP) EED,  a power-law log-parabola (PLLP) EED and a broken power-law (BPL) EED, are assumed in fits. We find that the one-zone SSC model with the PLLP EED provides successful fits to all the twelve SEDs. However, the one-zone SSC model with the LP and BPL EEDs fail to provide acceptable fits to the highest energy X-ray data or GeV data in several states. We therefore conclude that the one-zone SSC model works well in explaining the SEDs of Mrk 421, and the PLLP EED is preferred over the LP and BPL EEDs for Mrk 421 during the flare in March 2010. We derive magnetic field $B'\sim0.01$\ G, Doppler factor $\delta_{\rm D}\sim$30--50, and the curvature parameter of EED $r\sim1$--$10$ in the model with the PLLP EED. The evolutions of model parameters are explored. The physical implications of our results are discussed.
\end{abstract}


\section{Introduction}

Blazars are a class of radio-loud active galactic nuclei (AGN) whose jets point to Earth.
Blazar emission extends from MHz radio frequencies to TeV gamma-ray energies.
Their spectral energy distributions (SEDs) have two bumps, one peaking at infrared to X-rays,
and the other peaking in gamma-ray energies.
It is believed that the low-energy bump is the synchrotron emission from high-energy electrons in the jet.
The origin of the high-energy bump is still under debate.
Leptonic models and hadronic models have been proposed to explain the origin of the high-energy bump.
In the leptonic models, the high-energy bump is the inverse-Compton emission from high-energy electrons scattering low-energy
photons \citep[e.g.,][]{Dermer93,Sikora1994}.
In the hadronic models, the high-energy bump is the synchrotron emissions radiated by high-energy protons or secondary particles produced
in proton-photon interactions \citep[e.g.,][]{Mannheim,mucke2001,bottcher13}.

For high-synchrotron-peaked (HSP) BL Lac objects whose synchrotron peak
frequency is greater than $10^{15}\ $Hz \citep{Abdo10}, a one-zone leptonic synchrotron-self Compton (SSC) model usually
provides excellent fits to their SEDs \citep{Abdo11a,Abdo11b,Man,zhang12,Yan14,Zhou14}.
Mrk 421 and Mrk 501 are two of the closest (the redshift $z=0.031$ and $z=0.034$ respectively) and brightest TeV HSPs.
Many multiwavelength monitoring campaigns for the two typical HSPs have been organized to study
their broadband SEDs \citep[e.g.,][]{Abdo11b,Aleksi15,Balokovic,Bartoli,Furniss}.
We recently notice that new extensive broadband data seem to challenge the one-one SSC model for HSPs.
\citet{Shukla} constructed the simultaneous SEDs of Mrk 501 from observations during 2011.
They claimed that a one-zone SSC model cannot explain the SED of Mrk 501 during 2011 April-May, due to the hard {\it Fermi}-LAT spectrum.
\citet{Furniss} reported the simultaneous broadband observations of Mrk 501 between 2013 April
1 and August 10, including the first {\it NuSTAR} observations, and modelled the SEDs with a one-zone SSC model.
Looking at their modelling results, one can see that the one-zone SSC model cannot reproduce the highest TeV data.
\citet{Balokovic} presented the simultaneous broadband observations of Mrk 421 taken in
2013 January-March, and they modelled these SEDs with a one-zone SSC model.
One can see that the SSC model is inconsistent with the GeV-TeV spectrum obtained during
2013 January 15. Mrk 421 was observed at multiwavelengthes for 13 consecutive days during March 2010,
and its simultaneous SEDs with unprecedented
wavelength coverage from radio frequencies to GeV-TeV energies were built \citep{Aleksi15}.
\citet{Aleksi15} found that in several states the one-zone SSC model does not matches the observed data.
In order to match the data, \citet{Shukla} and \citet{Aleksi15} developed a two-zone SSC model for Mrk 501 and Mrk 421.

Although several examples that challenge the one-zone SSC model for Mrk 421 and Mrk 501 are summarized above,
one cannot exclude the one-zone SSC model for the two typical HSPs.
In our view a key point is missing, i.e., the above studies dot not perform searching for parameter space,
and therefore the modelling result may not be the best-fit result.

It is important to find convinced evidence for the failures of one-zone SSC model for HSPs.
It will motivate us to develop new models, and to find new emission mechanisms,
which has a big impact on our understandings of the blazar jet physics.

As mentioned above, \citet{Aleksi15} reported the day-scale SEDs of Mrk 421 during a flare state in March 2010.
These SEDs have unprecedented wavelength coverage.
We adopt these high-quality SEDs to test the one-zone SSC model.
Given that we do not know the electron energy distribution (EED) in the emission region, we assume three kinds of EED,
i.e., a log-parabola (LP) EED, a power-law log-parabola (PLLP) EED and a broken power-law (BPL) EED.
The Markov chain Monte Carlo (MCMC) technique is used to search high-dimension parameter space, and to obtain the best-fit result.
Throughout this paper, the cosmology with $H_0=71\rm \ km\ s^{-1}\ Mpc^{-3}$, $\Omega_{\rm m}=0.27$, and $\Omega_{\Lambda}=0.73$ is adopted.

\section{One-zone SSC models}

The one-zone model assumes that non-thermal radiations are produced in a single, homogeneous and spherical region in the jet.
The emission region moves relativistically toward us, and consequently the intrinsic radiation is strongly amplified due to the Doppler boosting.
Three parameters are needed to characterise the emission region: the comoving magnetic field $B'$, the Doppler factor $\delta_{\rm D}$,
and the comoving radius of the emission region $R'_{\rm b}$ which is expressed as $R'_{\rm b}=c\delta_{\rm D}t_{\rm var}/(1+z)$
where $t_{\rm var}$ is the minimum measured variability timescale.

High-energy electrons in the region produce synchrotron photons, and also up-scatter these synchrotron photons to higher energies.
The EED in the emission region is uncertain. We adopt three commonly used EEDs.
One is the PLLP EED \citep[e.g.,][]{Yan13,Zhou14,Dermer15},
\begin{equation}
\gamma'^2 N'_e\left({\gamma'}\right)\sim
\begin{cases}
\left( \frac{\gamma'}{\gamma'_{\rm c}}\right)^{2-s}&                                                     {\gamma'}\le{\gamma'_{\rm c}}\\
\left( \frac{\gamma'}{\gamma'_{\rm c}}\right)^{2-[ s+r\ {\rm log}\left(\frac{\gamma'}{\gamma'_{\rm c}}\right)]}&  {\gamma'}>{\gamma'_{\rm c}}
\end{cases}\ \ ,
\end{equation}
where $s$ is the power-law spectral index of the low-energy
branch, $\gamma'_{\rm c}$ is the cut-off energy of the power-law component,
and $r$ is a log-parabola width parameter.
The comoving electron energy density is written as $u'_{\rm e}=\zeta_{\rm e}u'_{\rm B}$,
where $u'_{\rm B}=B^{' 2}/(8\pi)$ is the comoving magnetic field energy density, which is used to normalize the EED.
Another one is the LP EED \citep{Dermer14,Dermer15,Yan151},
\begin{equation}
\gamma'^2 N'_e(\gamma') \sim\left( \frac{\gamma'}{\gamma'_{\rm pk}} \right)^{-b\ \log{\left(\frac{\gamma'}{\gamma'_{\rm pk}}\right)}}\ ,
\end{equation}
 where $b$ is the spectral curvature parameter, and $\gamma'_{\rm pk}$ is the peak Lorentz factor in the $\gamma'^2 N'_e(\gamma')$ distribution.
 The third one is the broken power-law (BPL) EED \citep{Finke08}
 \begin{eqnarray}
\gamma'^2 N_{\rm e}^{\prime}(\gamma^{\prime})\sim H(\gamma^{\prime};\gamma_{\rm min}^{\prime},\gamma_{\rm
max}^{\prime})\{{\gamma^{\prime
2-p_1}\exp(-\gamma^{\prime}/\gamma_{\rm b}^{\prime})}
\nonumber \\
\times H[(p_{\rm 2}-p_{\rm 1})\gamma_{\rm
b}^{\prime}-\gamma^{\prime}]+[(p_{\rm 2}-p_{\rm 1})\gamma_{\rm
b}^{\prime}]^{p_{\rm 2}-p_{\rm 1}}\gamma^{\prime 2-p_{\rm 2}}
\nonumber \\
\times \exp(p_{\rm 1}-p_{\rm 2})H[\gamma^{\prime}-(p_{\rm
2}-p_{\rm 1})\gamma_{\rm b}^{\prime}]\},
\end{eqnarray}
where $H(x;x_{1},x_{2})$ is the Heaviside function:
$H(x;x_{1},x_{2})=1$ for $x_{1}\leq x\leq x_{2}$, and
$H(x;x_{1},x_{2})=0$ everywhere else; as well as $H(x)=0$ for
$x<0$ and $H(x)=1$ for $x\geq0$. $\gamma_{\rm min}^{\prime}$ and
$\gamma_{\rm max}^{\prime}$ are the minimum and maximum energies
of electrons, respectively. The spectrum is
smoothly connected with indices $ p_1$ and $p_2$ below and above
the break energy $\gamma_{\rm b}^{\prime}$.

The three EEDs have clear physical origins. An initial single power-law EED
can be deformed to become a BPL due to energy losses of electrons \citep[e.g.,][]{Yan162}.
LP function is motivated by the second-order Fermi acceleration \citep[e.g.,][]{Becker06}.
Considering a power-law distribution of electrons injected into a turbulent region
where second-order processes broaden the distribution, the EED approximates the PLLP function \citep{Dermer15}.

 The SSC radiation spectrum is calculated by the method given by \citet{Finke08}.
 We adopt the extragalactic infrared-UV background light (EBL) model of \citet{Finke10} to correct the absorption effect.
 This EBL model is consistent with other popular EBL models  \citep[e.g.,][]{Franceschini,Domi,Inoue}.

 The MCMC technique based on the Bayesian statics is a very powerful fitting tool.
  It is well suitable to search high-dimension parameter space, and to obtain the uncertainties of the model parameters. Our MCMC code is adapted from the COSMOMC package\footnote{http://cosmologist.info/cosmomc/} \citep{Lewis} by \citet{liu12}. \citet{Fan} and \citet{yuan11} used this code to fit the SEDs of supernova remnants.
 \citet{Yan13} applied the MCMC technique to the SEDs of blazars \citep[also see e.g.,][]{Zhou14,Dermer15,Yan16}.

\begin{figure*}
	   \centering
		\includegraphics[width=505pt,height=155pt]{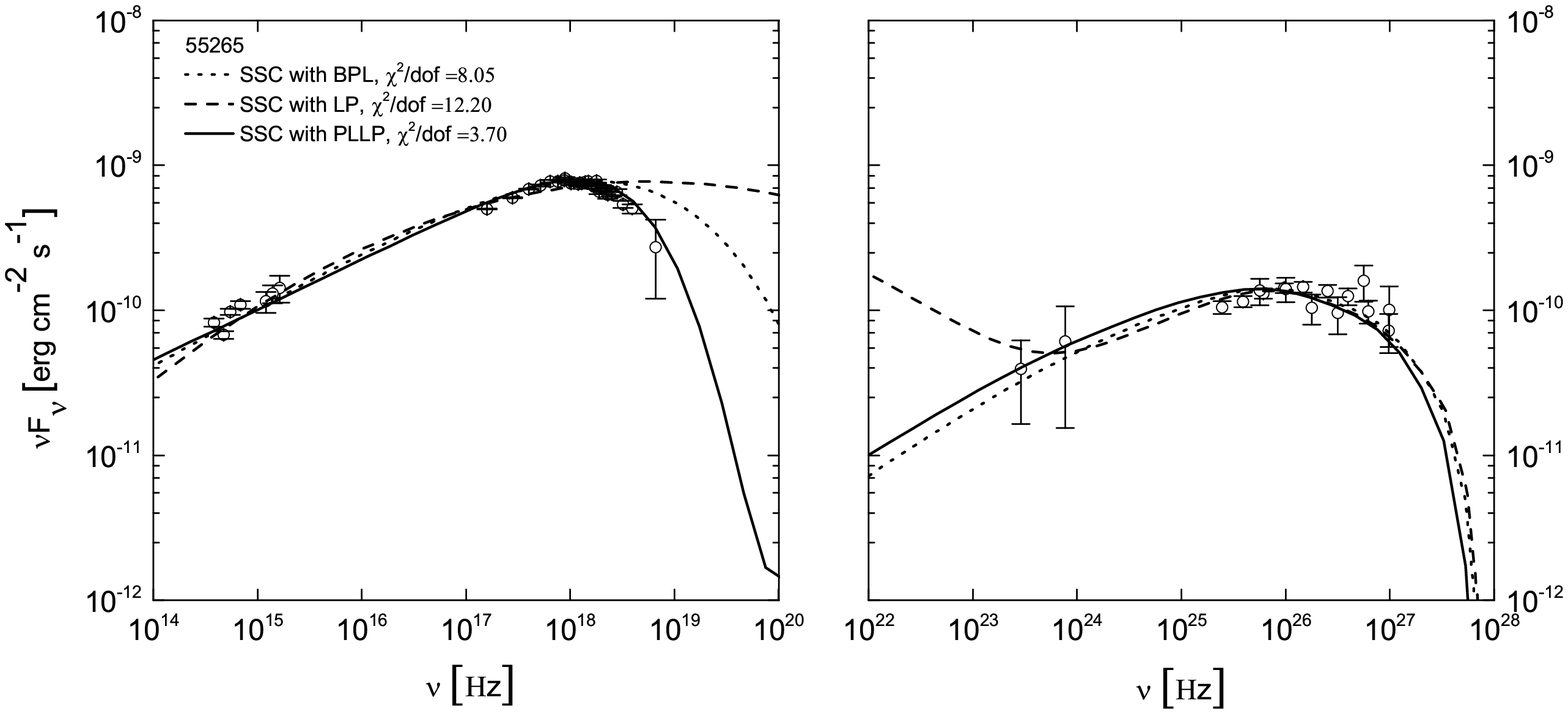}
		\includegraphics[width=505pt,height=155pt]{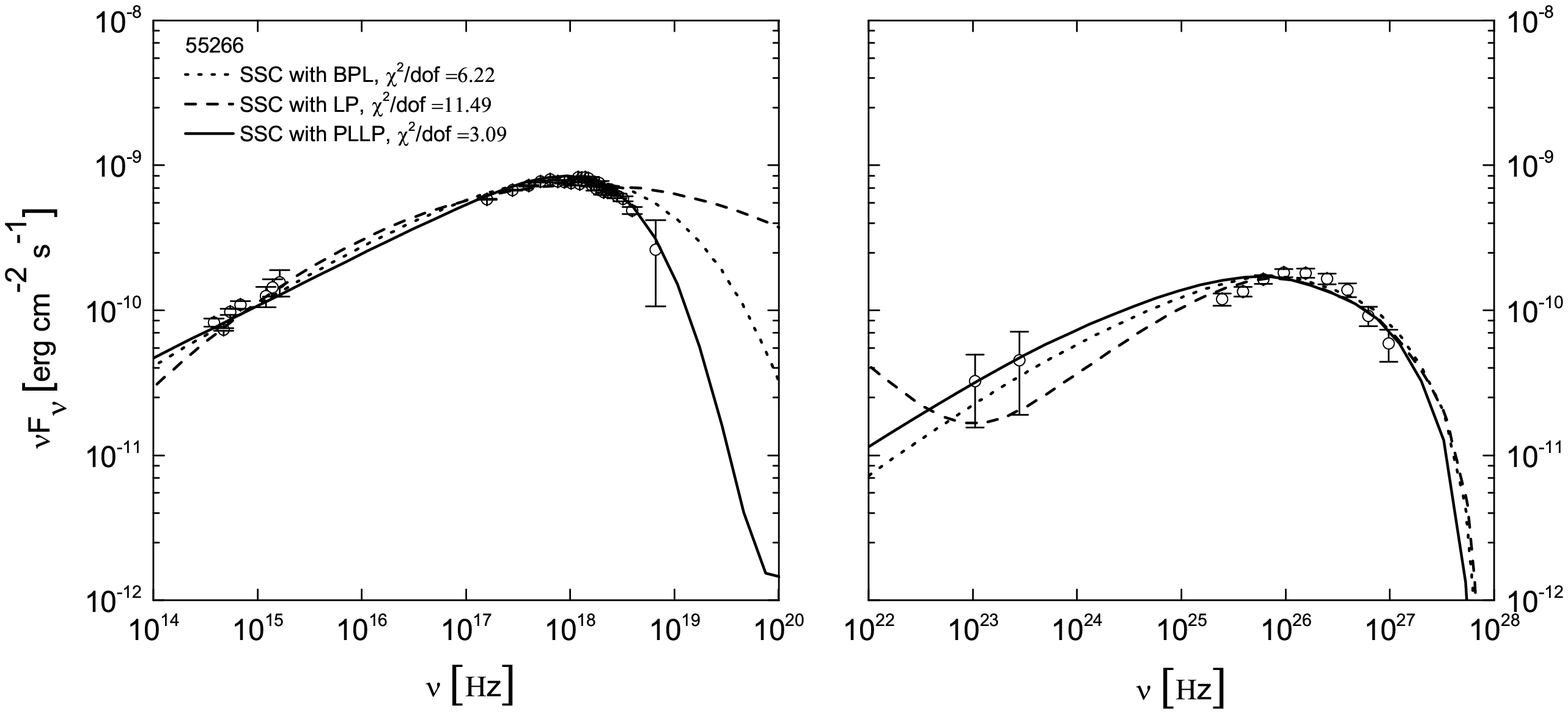}
		\includegraphics[width=505pt,height=155pt]{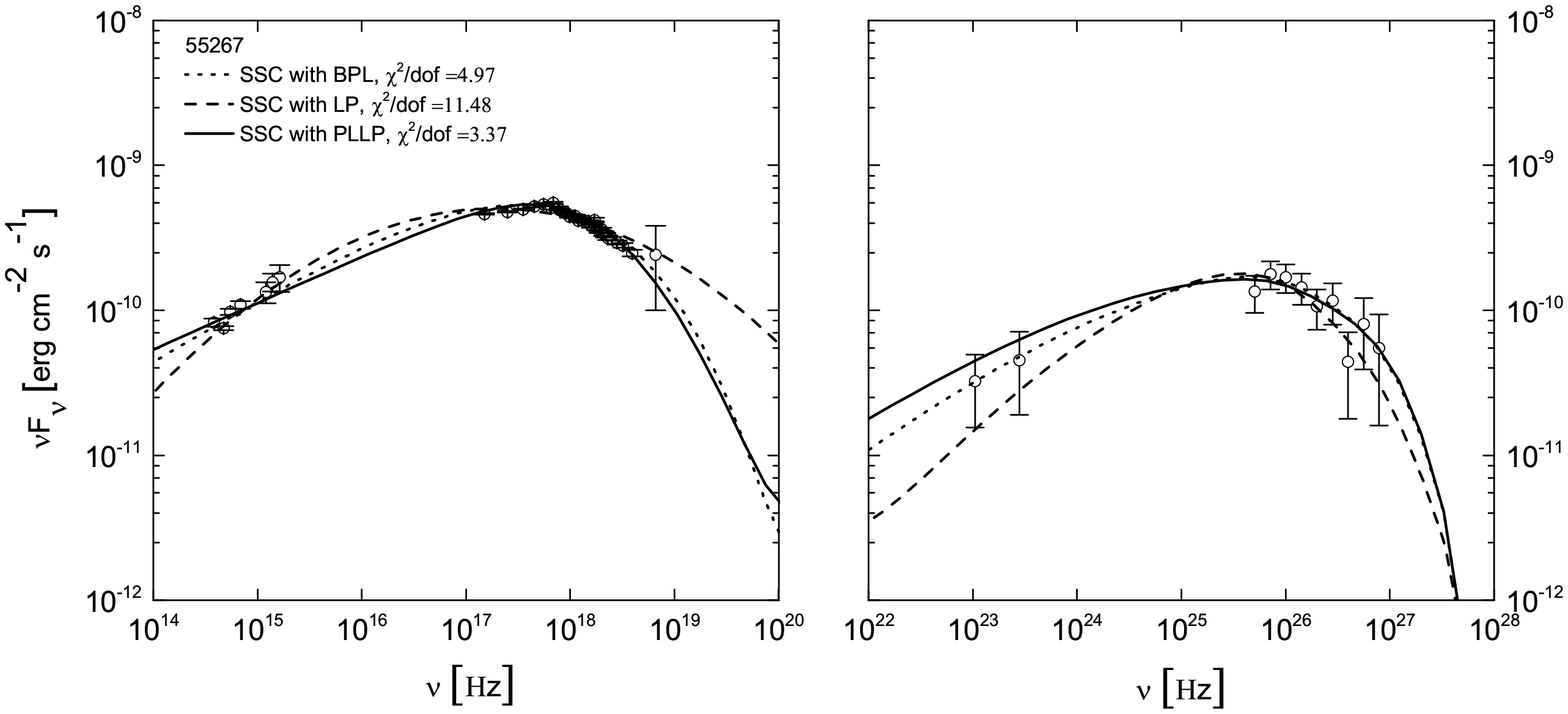}
        \includegraphics[width=505pt,height=155pt]{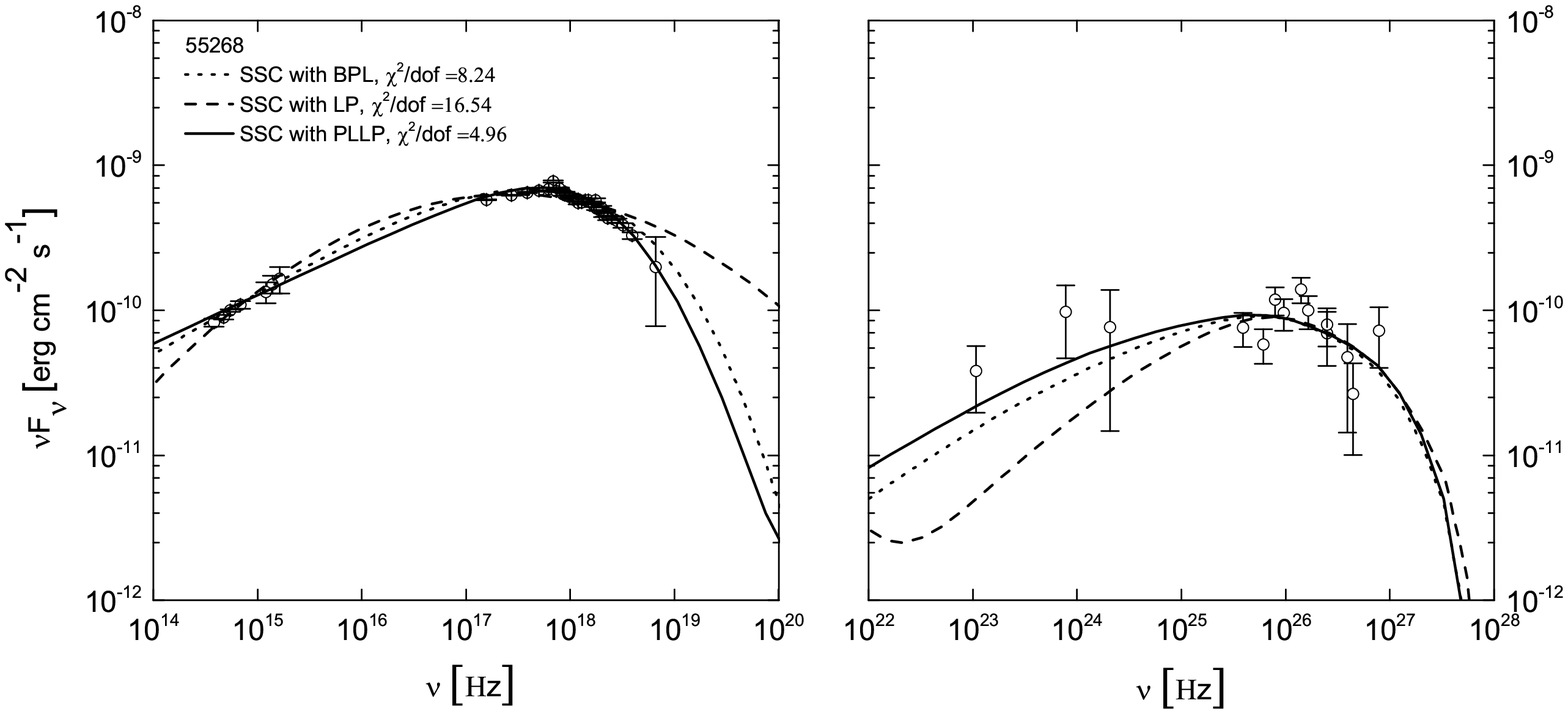}
\caption{The best fits to the SEDs of Mrk 421 during MJD 55265, 55266, 55267, and 55268 \citep[data from][]{Aleksi15}. Solid line: the synchrotron/SSC spectrum with the PLLP EED; Dashed line: the synchrotron/SSC spectrum with the LP EED; Dotted line: the synchrotron/SSC spectrum with the BPL EED. The reduced $\chi^2_{\rm red}$=$\chi^2$/dof derived in the fits are reported. Left: low-energy bump; right: high-energy bump. Note that though the two bumps in each SED are shown separately the fit is performed simultaneously. \label{sed1}}
\end{figure*}

\begin{figure*}
	   \centering
		\includegraphics[width=505pt,height=155pt]{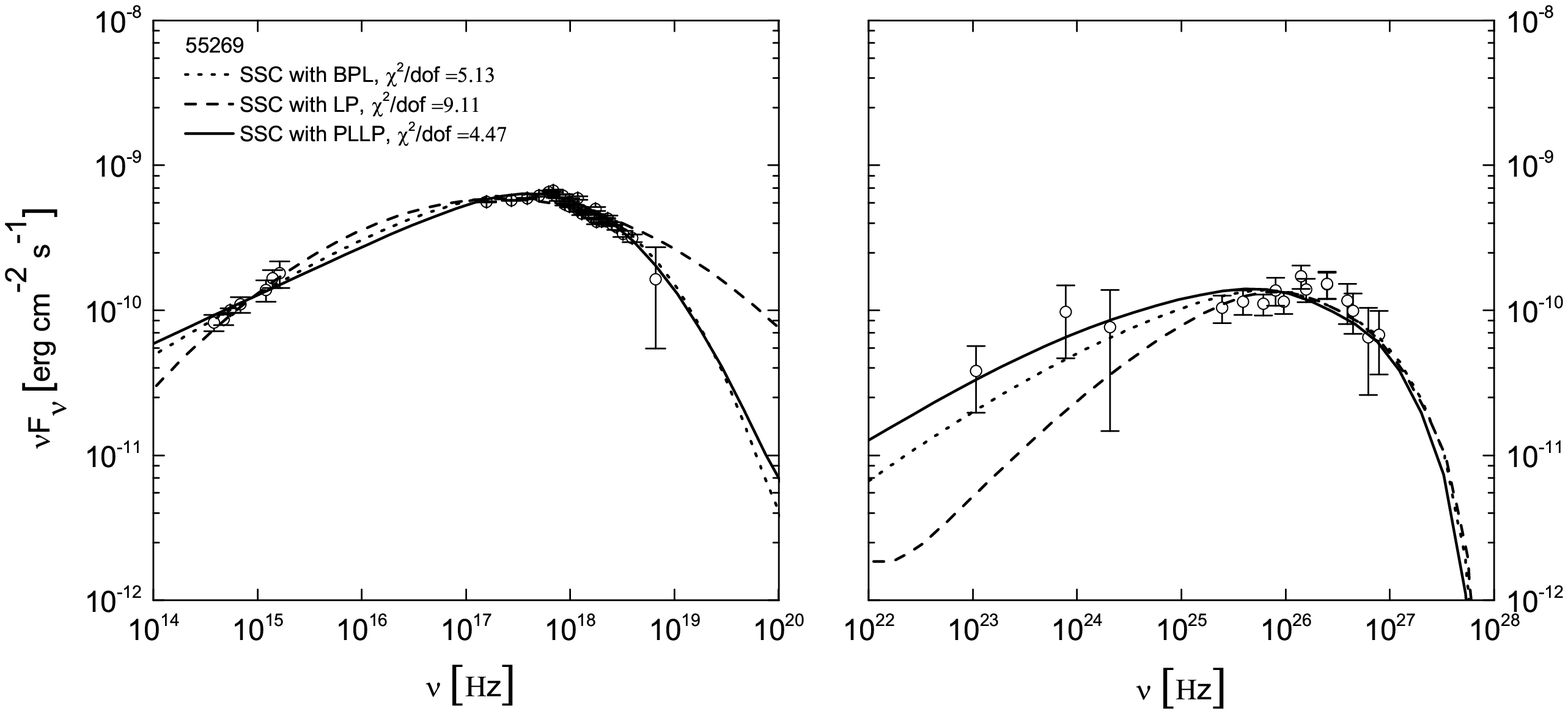}
		\includegraphics[width=505pt,height=155pt]{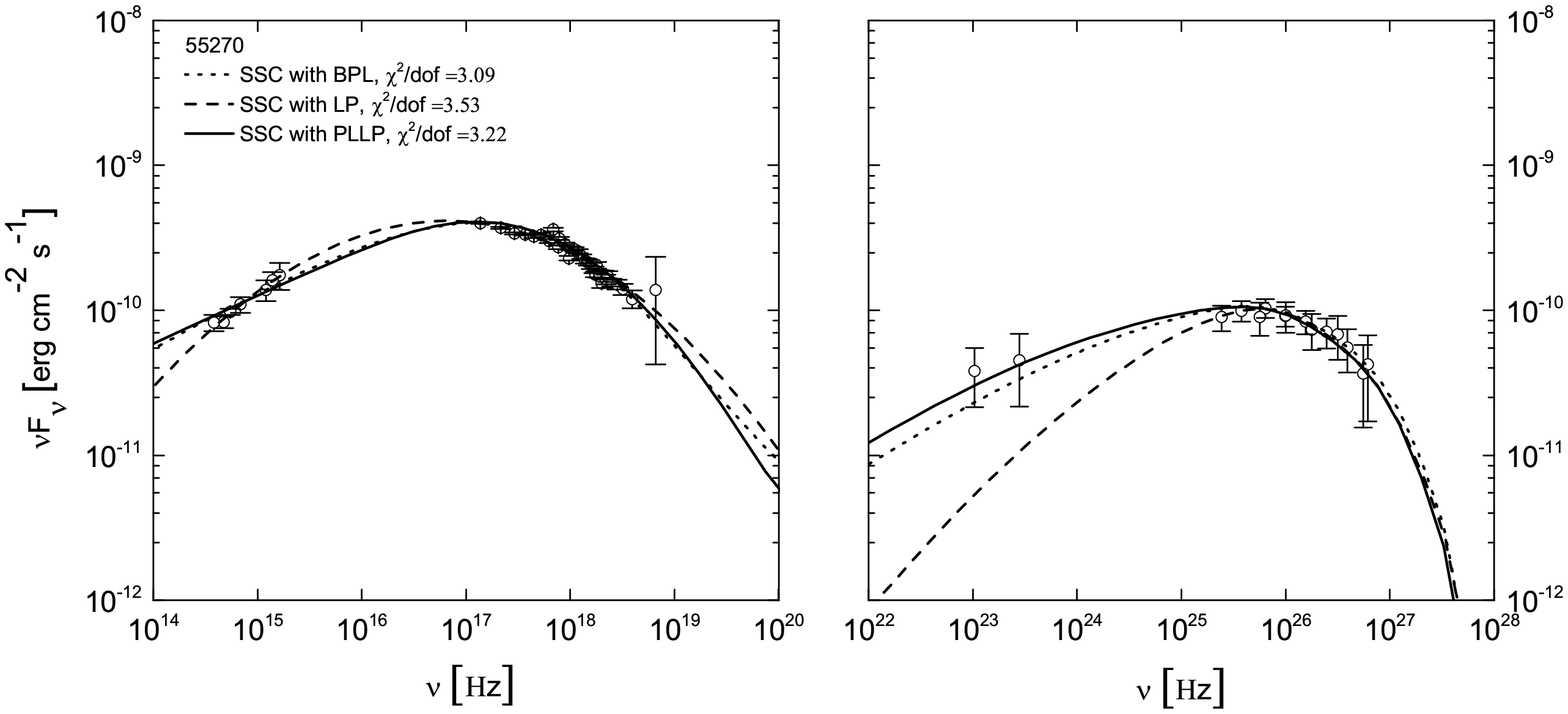}
        \includegraphics[width=505pt,height=155pt]{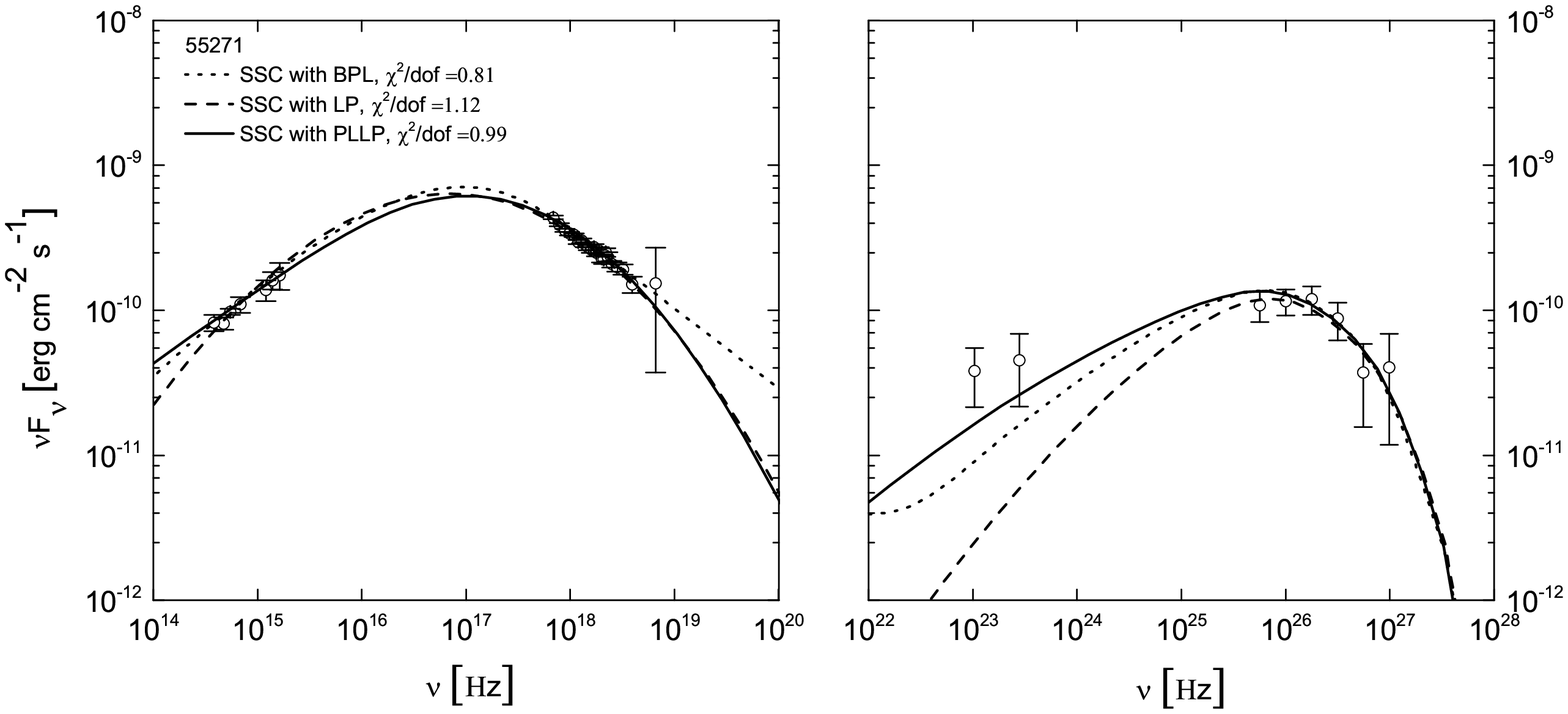}
		\includegraphics[width=505pt,height=155pt]{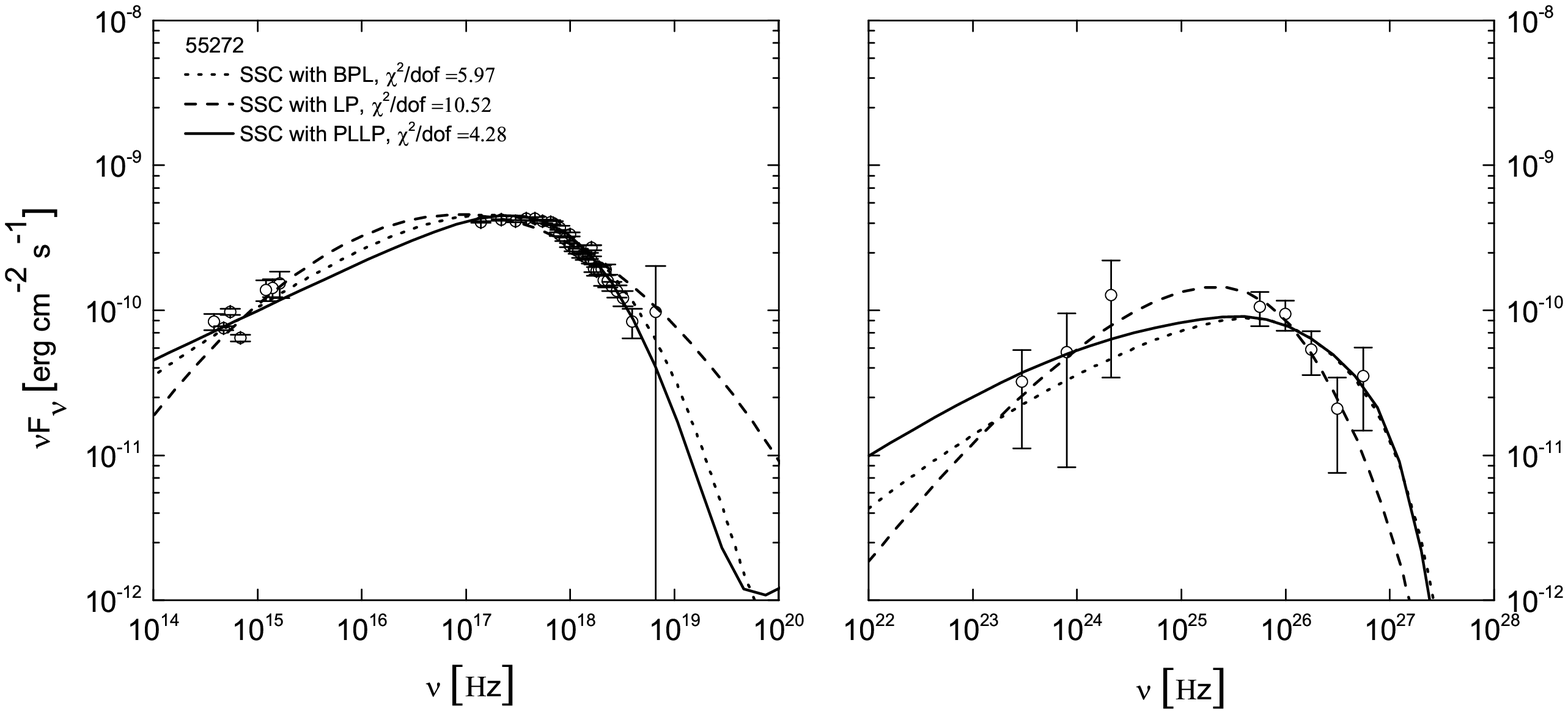}
\caption{Same as Fig.~\ref{sed1}, but for the SEDs during MJD 55269, 55270, 55271, and 55272.\label{sed2}}
\end{figure*}

\begin{figure*}
	   \centering
	  \includegraphics[width=505pt,height=155pt]{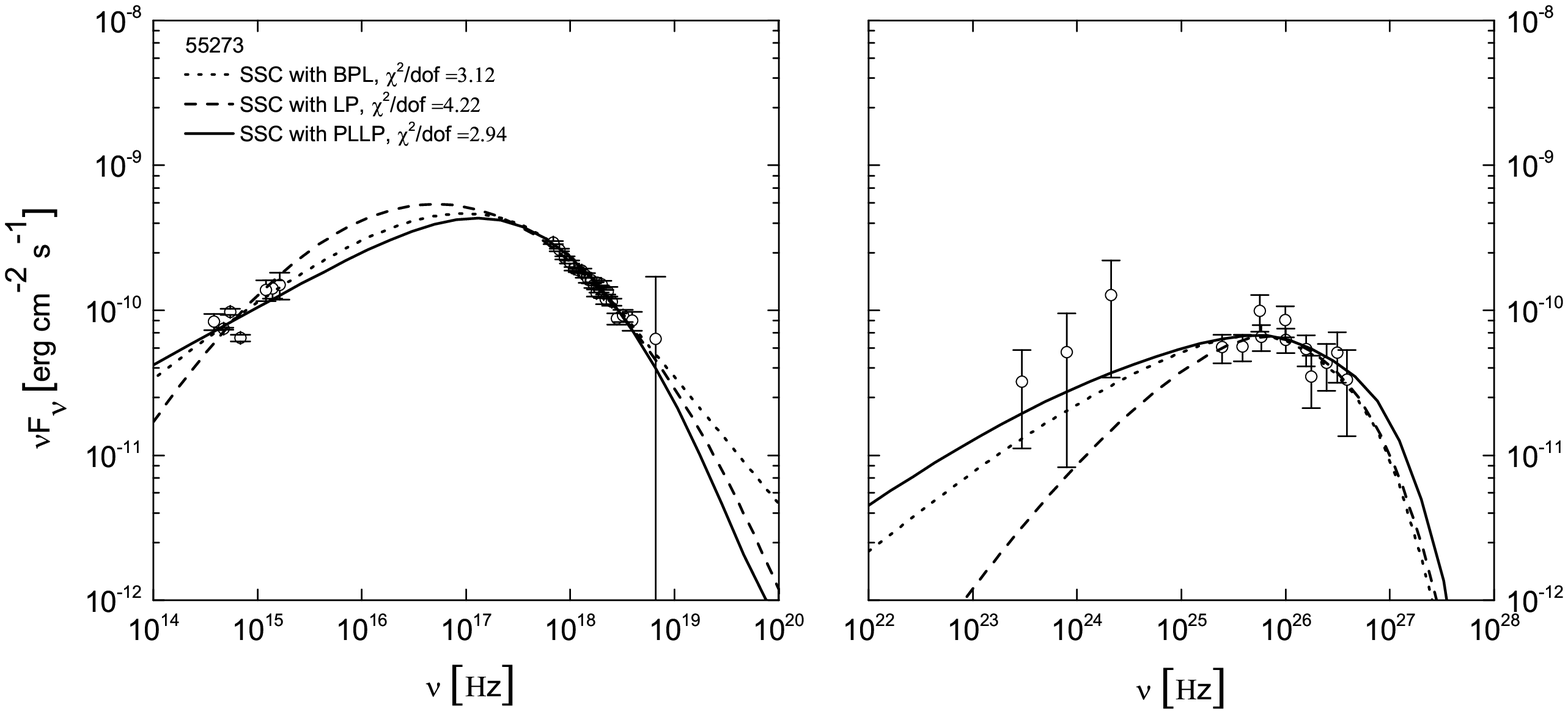}
	  \includegraphics[width=505pt,height=155pt]{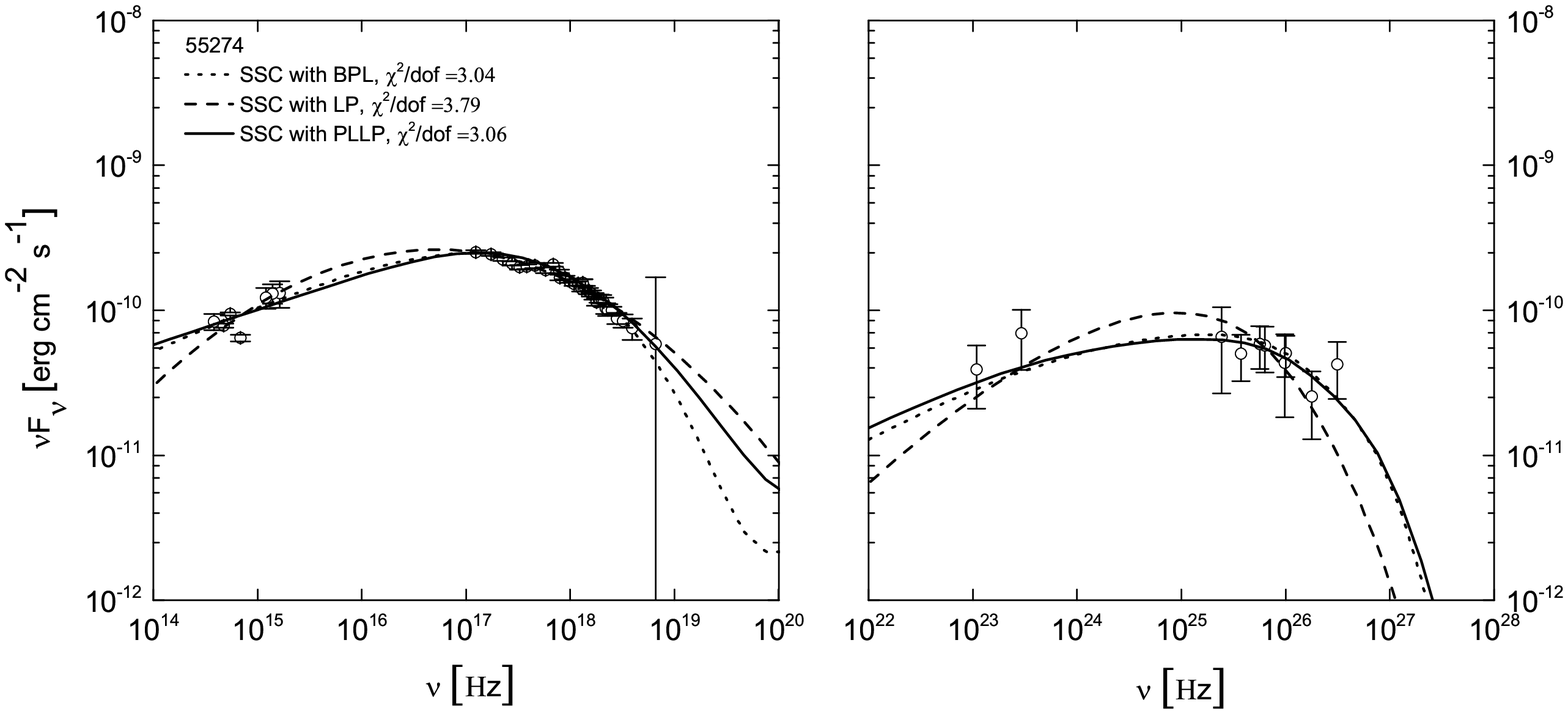}
	  \includegraphics[width=505pt,height=155pt]{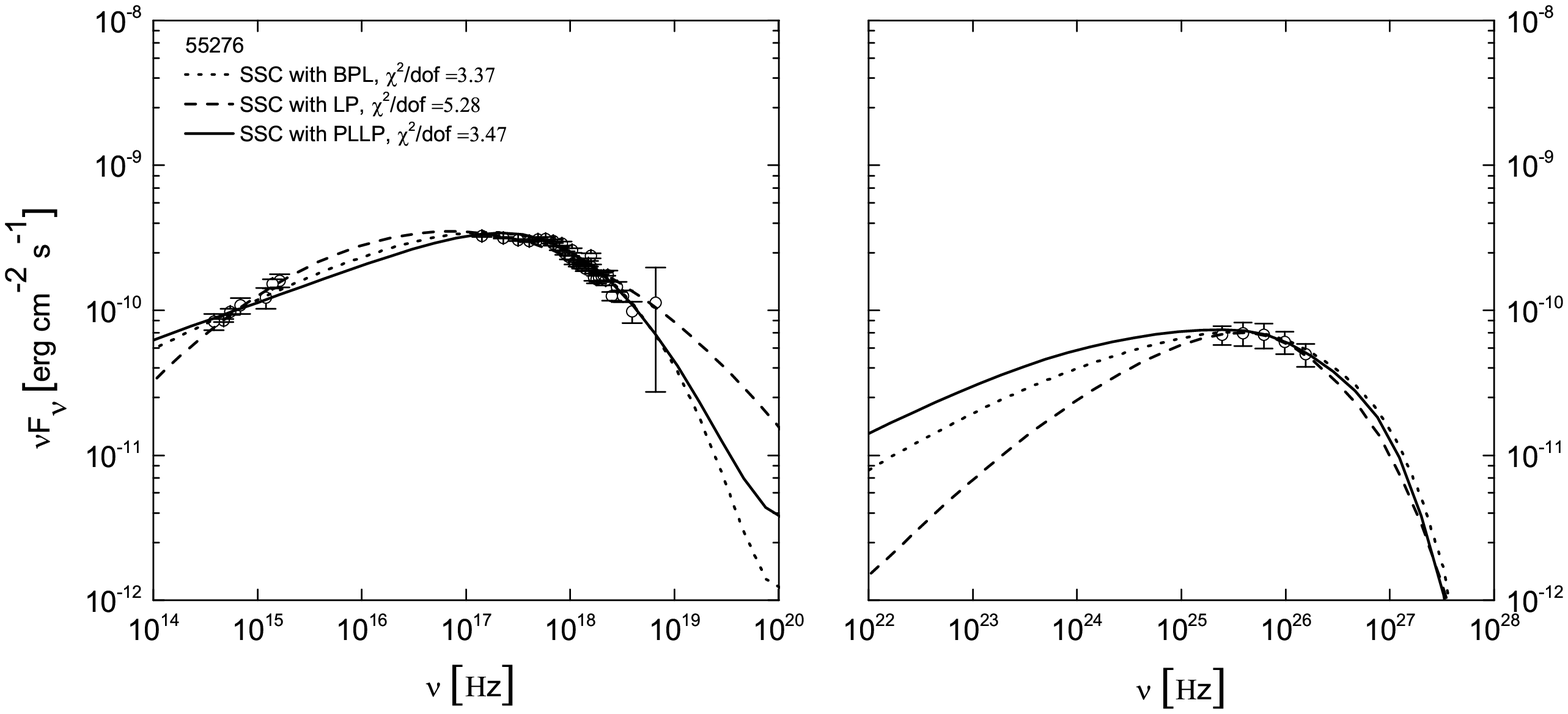}
	  \includegraphics[width=505pt,height=155pt]{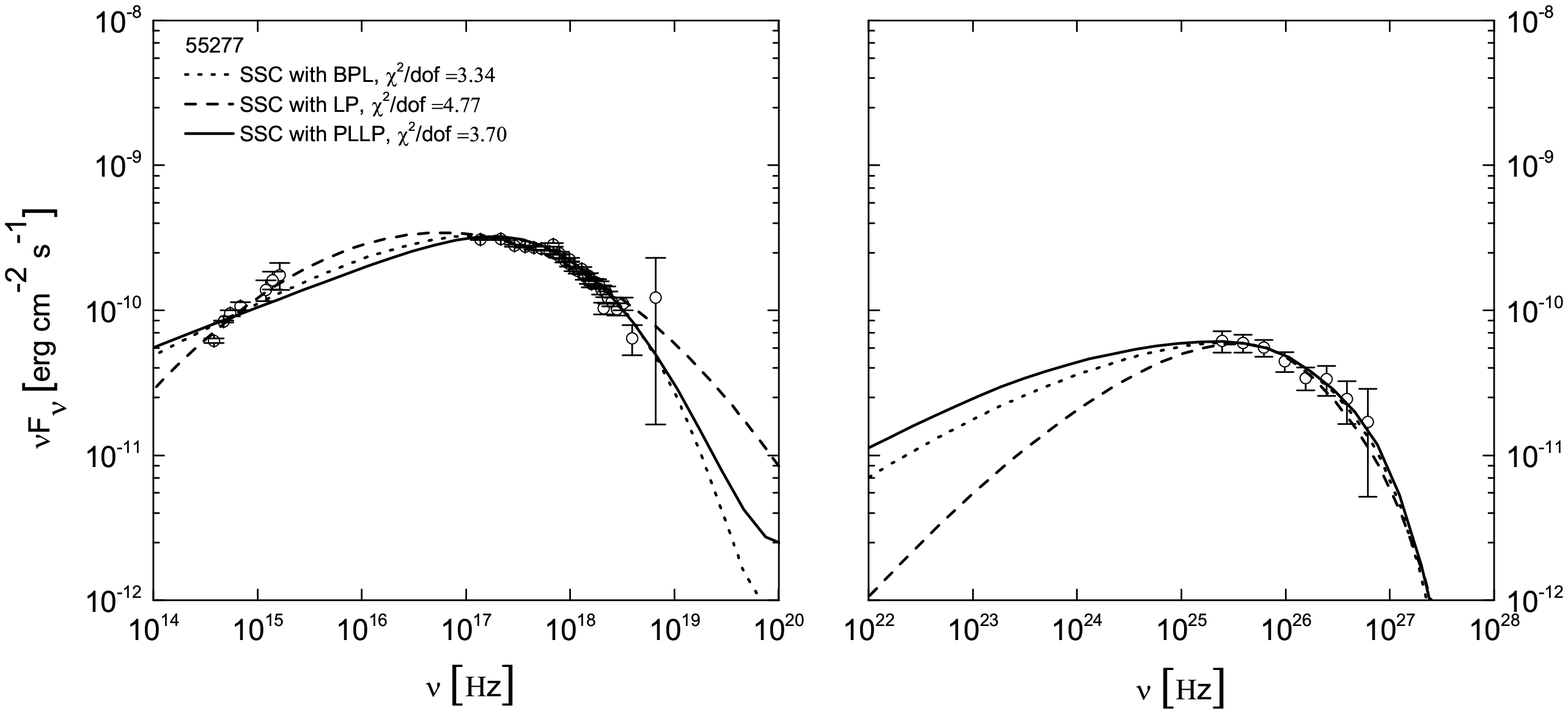}	
\caption{Same as Fig.~\ref{sed1}, but for the SEDs during MJD 55273, 55274, 55276, and 55277.\label{sed3}}
\end{figure*}

\begin{table*}
\begin{center}
 \caption{Model parameters derived in the one-zone SSC model with the PLLP EED. The mean values and the marginalized 68\% confidence intervals (CI) are reported.}
\label{input2}
\begin{tabular}{@{}ccccccccc}
\hline
MJD &    $\zeta_e$   & $r$   & $B'$       & $s$  & $t_{\rm var}$   & $\delta_{\rm D}$   & $\gamma'_{\rm c}$   & $\nu^{\rm pk}_{\rm s}$ \\
&                                 &           &(0.1 G)                      &       &   ($10^4\ $s)        &                  &($\rm 10^{5}$)  & ($10^{17}\ $Hz) \\
\hline
55265         & 174        & 8.97            & 0.092            & 2.31     & 8.0     &32.9        &11.8    & 15 \\
 (68\% CI)       &167-200     &8.75-10    & 0.085-0.100   & 2.30-2.32  &-       &31.8-33.9    &11.4-12.2  & 14-16\\
 \hline
55266         & 173          & 7.54       & 0.098              &  2.28         & 8.0       &31.9          & 9.87    & 11\\
 (68\% CI)       & 169-200   & 6.75-10  & 0.089-0.11     &  2.27-2.29    &-          & 30.8-32.8    &9.18-10.53  & 10-12\\
\hline
55267         & 145           & 2.37       & 0.13              &  2.35         & 8.0     &28.7          & 4.45    & 2.7\\
 (68\% CI)       & 127-200   & 2.18-2.57  & 0.01-0.14     &  2.34-2.37     & -         & 27.7-29.3   &4.12-4.77 & 2.4-2.9\\
 \hline
55268         & 128            & 3.19       & 0.12               &  2.33         & 3.0     &50.0        &4.41     & 4.0\\
 (68\% CI)       & 100-199    & 3.00-3.30  & 0.10-0.13     &  2.32-2.34     & -      &47.0-52.4   &3.94-4.84   & 3.6-4.3\\
 \hline
55269         & 159           & 1.98       & 0.09               &  2.33         & 8.0    &34.4         &4.31     & 2.2\\
 (68\% CI)       & 147-200    & 1.80-2.17  & 0.01-0.10     & 2.32-2.34     & -       &33.7-35.4     &3.99-4.62  & 1.9-2.4\\
 \hline
55270         & 98          & 1.09        & 0.126              &  2.30         & 8.0    &31.6           &1.50     & 0.3\\
 (68\% CI)       & 89-113    & 0.98-1.19  & 0.001-0.13     & 2.26-2.35     & -      &31.1-32.9    &1.21-1.78   & 0.2-0.4\\
 \hline
55271         & 85            & 1.22       & 0.099              &  1.99         & 8.0    & 37.7          & 1.27     & 0.2\\
 (68\% CI)       & 63-100    & 1.06-1.38   & 0.001-0.117     & 1.80-2.18     & -        & 33.1-42.4       &0.85-1.76   & 0.04-0.4 \\
 \hline
55272         & 110         & 4.11      & 0.11               &  2.32         & 8.0      &33.8           &4.38    & 2.3\\
 (68\% CI)       & 80-200     &3.55-4.68  & 0.001-0.12    & 2.30-2.33     &  -     &31.4-35.8    &3.80-4.93   &2.1-2.6\\
 \hline
 55273        & 53            & 2.18      & 0.18             &  2.25         & 8.0    &30.4          &2.13      &0.8\\
 (68\% CI)       &1-78       & 1.84-2.57  & 0.10-0.22     & 2.16-2.34     & -     &26.9-34.5     &1.90-2.43   &0.6-0.9\\
 \hline
55274         & 87.8             & 1.21       & 0.22               &  2.52         & 8.0    &26.9     &1.88     & 0.7\\
 (68\% CI)       & 40-120       & 1.04-1.37  & 0.13-0.29     &  2.49-2.55     & -     &23.1-30.0  &1.55-2.14  & 0.4-0.9\\
 \hline
55276         & 134            & 1.89       & 0.11              &  2.48        & 8.0      &33.4      & 3.23     &1.4\\
 (68\% CI)       & 112-200      & 1.66-2.12  & 0.001-0.12     &  2.47-2.50    & -     &32.1-35.3  &2.83-3.59   & 1.2-1.7\\
 \hline
55277          & 127            & 1.90      & 0.112             &  2.45        & 8.0     &34.8        &2.97      & 1.1\\
 (68\% CI)       &97-200        & 1.71-2.10   & 0.001-0.113     & 2.43-2.46    & -     &33.6-37.5     &2.44-3.41   & 0.9-1.3\\
 \hline
\end{tabular}
\end{center}
\end{table*}

\section{Results}

\cite{Aleksi15} presented 13 day-scale SEDs of Mrk 421.
The details on the data and the telescopes can be found in \cite{Aleksi15}.
The SED during MJD 55275 has no simultaneous TeV data, therefore we do not include this SED in our study.
Using the MCMC technique we fit the rest of 12 SEDs with the one-zone SSC models with three types of EED.
The minimum variability timescale for each SED is determined by the observations in \cite{Aleksi15}, which is fixed in the fit.

\begin{figure}
	  \includegraphics[width=260pt,height=200pt]{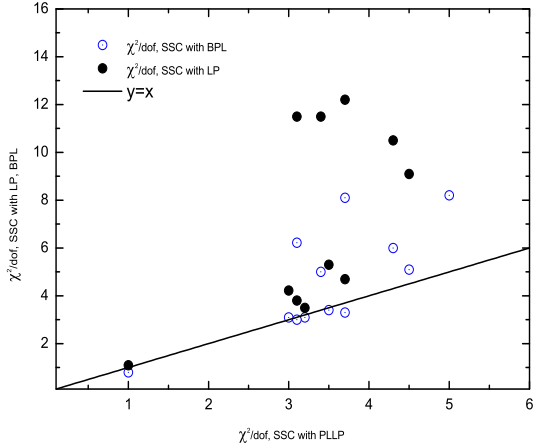}
\caption{Comparing the reduced $\chi_{\rm red}^2$ derived in the fits to the 12 SEDs with the three EEDs. \label{chi2}}
\end{figure}

Figs.~\ref{sed1}-\ref{sed3} show the best-fits to the 12 SEDs.
It can be easily seen that the one-zone SSC model with the PLLP EED provide excellent fits to all the 12 SEDs;
while the one-zone SSC model with the LP and BPL EEDs cannot provide acceptable fits to the highest energy X-ray data in MJD 55265 and MJD 55266.
The SSC model with LP EED predicts very hard GeV spectrum in each state and underestimates the GeV emissions.
The LP-EED model fails to fit the highest energy X-ray data or the GeV data in most states.
To systematically compare the goodness-of-fits with the three EEDs,
we plot the reduced $\chi^2_{\rm red}$=$\chi^2$/dof derived in the fits with the three EEDs in Fig.~\ref{chi2}.
It can been seen that overall the SSC model with the PLLP EED provide the best fits to the 12 SEDs.
In six states, the fits with the BPL EED are comparable to the fits with the PLLP EED,
while in the rest of six states the fits with the BPL EED are significantly worse than the fits with the PLLP EED.
It is clear that the PLLP EED is preferred over the LP and BPL EEDs during the flare state in March 2010.

\begin{figure}
	  \includegraphics[width=260pt,height=200pt]{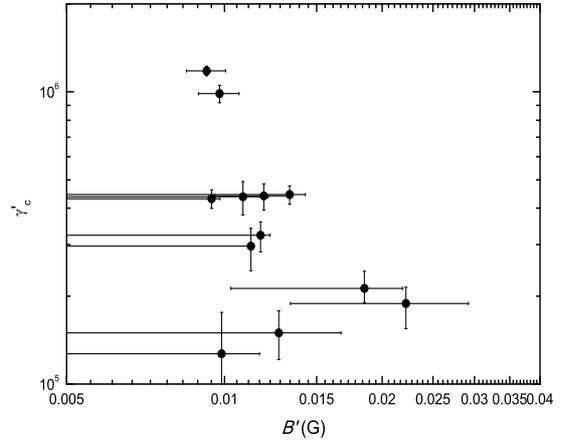}
\caption{The magnetic field $B'$
versus the cut-off energy of PLLP EED $\g'_{\rm c}$. \label{g-c}}
\end{figure}

\begin{figure}
	  \includegraphics[width=260pt,height=200pt]{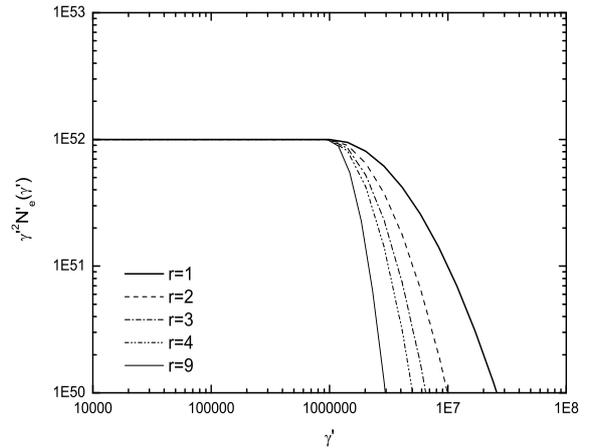}
\caption{An example to show the impact of the parameter $r$ on the high-energy cut-off shape of PLLP EED. \label{eed}}
\end{figure}

\begin{figure}
	  \includegraphics[width=260pt,height=200pt]{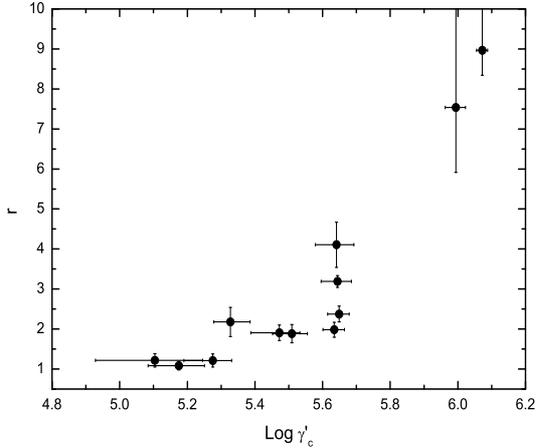}
\caption{The curvature parameter of PLLP EED $r$
versus the cut-off energy $\g'_{\rm c}$. \label{rg}}
\end{figure}

In Table~\ref{input2} we list the model parameters derived in the one-zone SSC with the PLLP EED.
In Appendix~\ref{app}, we give the model parameters derived in the one-zone SSC with the LP and BPL EEDs (Table~\ref{inputlp} and Table~\ref{inputbpl}). Here, we focus on analyzing the model parameters derived with
the PLLP EED. In Fig.~\ref{g-c}, we show the $B'$--$\g'_{\rm c}$ plot.
One can see that the magnetic field $B'$ \textbf{is} $\sim0.01\ $G during these days.
The electron cut-off energy $\g'_{\rm c}$ varies from $10^5$ to $10^6$.
It is noted that we derive $\g'_{\rm c}\approx10^6$ in MJD 55265 and MJD 55266, and a very large $r>7$.
In Fig.~\ref{eed}, one can see that such a large $r$ results in a very sharp high-energy cut-off, and the EED approximates a single power-law.
In the two sates (MJD 55265 and MJD 55266), we found the peak synchrotron frequency $\nu^{\rm pk}_{\rm s}\gtrsim10^{18}\ $Hz.
Such a large $\nu^{\rm pk}_{\rm s}$ is rare for HSPs \citep[see the peak synchrotron frequencies reported in][]{Tramacere11,Dermer15}.
Fig.~\ref{rg} shows the $r$--$\g'_{\rm c}$ plot. It seems that $\g'_{\rm c}$ is proportional to the curvature $r$.
We adopt a linear function to fit the data, and derive $r=(3.1\pm0.3)\ {\rm Log}\ \g'_{\rm c}-(15\pm1)$ with an adjusted $R^2=0.7$ and a chance probability $p=2.5\times10^{-4}$.

In MJD 55268 the TeV emission showed potential a intra-night variability \citep{Aleksi15}, hence we adopt $t_{\rm var}\thickapprox8\ $hr to fit the SED, and derive $\delta_{\rm D}\sim50$.
Except this state, we adopt $t_{\rm var}\thickapprox22\ $hr, and derive $\delta_{\rm D}\sim$30--35.

\begin{figure}
	  \includegraphics[width=260pt,height=200pt]{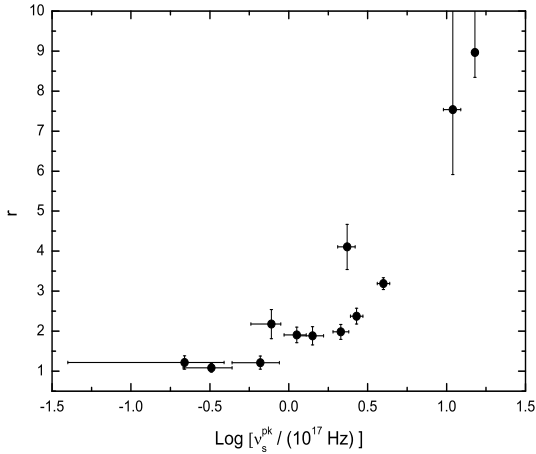}
\caption{The curvature parameter of PLLP EED $r$ versus the peak synchrotron frequency $\nu^{\rm pk}_{\rm s}$. \label{rv}}
\end{figure}

In Table~\ref{input2}, we also give the values of $\nu^{\rm pk}_{\rm s}$ evaluated by the model parameters
through the relation $\nu^{\rm pk}_{\rm s}\simeq3.6\times10^6\g'^{ 2}_{\rm c}B'\delta_{\rm D}\ $Hz.
In Fig.~\ref{rv}, we show the relation of $\nu^{\rm pk}_{\rm s}$ and $r$.
We define $\nu_{17}=\nu^{\rm pk}_{\rm s}/10^{17}$.
We also adopt a linear function to fit the data, and derive $r=(1.6\pm0.1)\ {\rm Log}\ \nu_{17}-(1.9\pm0.1)$ with an adjusted $R^2=0.8$ and a chance probability $p=6.1\times10^{-5}$.
Because of the relation $b_{\rm s}\simeq r/5$\footnote{Note that the relation $b_{\rm s}\simeq r/5$ is given by synchrotron emission theory \citep{Massaro4}, which is robust.} where $b_{\rm s}$ is the curvature of synchrotron bump \citep{Massaro4,Tramacere11},
the trend of $\nu^{\rm pk}_{\rm s}$--$r/b_{\rm s}$ is different from that reported in previous studies \citep[e.g.,][]{Tramacere09}
 where an inverse trend of $\nu^{\rm pk}_{\rm s}$--$b_{\rm s}$ is presented.

\section{Discussion AND Conclusions}

Taking advantage of the MCMC technique \citep{Yan13,Yan151},
we test the one-zone SSC models for Mrk 421 using the high-quality SEDs with unprecedented data coverage reported in \citet{Aleksi15}.
According to the fitting results, we conclude that the one-zone SSC model still works well in explaining the broadband SEDs.
There is no evidence of a second leptonic/hadronic emission component.
Furthermore, our study rules out the LP and BPL EEDs for Mrk 421, and supports the PLLP EED.
We determine the magnetic field $B'\sim0.01\ $G and the radius of blob $R'_{\rm b}\sim[5-8]\times10^{16}\ $cm in the PLLP model.

The curvature in electron distribution is related to second-order Fermi acceleration theory \citep[e.g.,][]{Becker06,Massaro6,sp08,Tramacere11,Yan12,Asano14}.
Second-order processes broaden the injected electron distribution,
and introduce a curvature into the energy distribution.
Our results that the EED in the jet of Mrk 421 is the PLLP EED implies a scenario
combining the first- and second-order Fermi acceleration processes, in which a power-law
distribution of particles injected downstream of a shock
into a turbulent region where the second-order Fermi acceleration processes broaden
the distribution, and then the PLLP EED is formed.
The curvature of PLLP EED $r$ in MJD 55265 and MJD 55266 is extremely large,
so that the EED is very close to a single power-law distribution.
This may be due to the cooling effect \citep{Tramacere11}.

The evolution of model parameters can reveal the information of physical processes \citep[e.g.,][]{Yan13,Yan16}.
We find that the cut-off energy $\g'_{\rm c}$ in the PLLP EED increases with the EED's curvature $r$.
\citet{Tramacere11} have shown that the curvature $r$ is inversely proportional
to the momentum diffusion coefficient when acceleration is dominated over cooling;
and $r$ quickly increases once the cooling becomes relevant.
The evolution of $\g'_{\rm c}$ with $r$ (Fig.~\ref{g-c}) hints that the radiative cooling of electrons become relevant and the EED approaches the equilibrium between acceleration and cooling.
Moreover, the trend of $\nu^{\rm pk}_{\rm s} - b_{\rm s}$ we derived (Fig.~\ref{rv}; using the relation $b_{\rm s}\simeq r/5$) is different from
the inverse correlation found in optical-X-ray data analysis on HSPs \citep[e.g.,][]{Massaro6,MassaroE,Tramacere07,Tramacere09}.
The inverse correlation between $\nu^{\rm pk}_{\rm s}$ and $b_{\rm s}$ may imply an acceleration-dominated scenario \citep{Tramacere11}.
From our resutls, one can see that the evolution of $\nu^{\rm pk}_{\rm s} - b_{\rm s}$ is the direct representation of the evolution of $\g'_{\rm c} - r$ in the observable space.

\citet{Yan13} analyzed two SEDs of Mrk 421, respectively, in a
quiescent state \citep{Abdo11b} and in a giant TeV flare \citep{Shukla12}, and found that the EED in the TeV flare is the PLLP.
The 12 SEDs analyzed in this work are obtained in a X-ray and TeV flare state in March 2010 \citep{Aleksi15}.
It seems that the PLLP EED that involves the first- and second-order Fermi acceleration processes works in the flare state of Mrk 421,
and the cooling timescale of electrons with $\g'_{\rm c}$ may be close to the acceleration timescale.
The physical mechanism in quiescent states of Mrk 421 is worth a systematical investigation in a separate study. 

As a last remark, we would like to note that an alternative model to explain the SEDs is the leptonic-hadronic model.
\citet{Petropouloua} explained the 13 SEDs in \citet{Aleksi15} well with a one-zone leptonic-hadronic model.
They derived $B'=5\ $G and $\delta_{\rm D}\sim20$, which are different from those derived in the leptonic model.
The hadronic model is attractive, since it predicts high-energy neutrinos.
To distinguish the leptonic model from the hadronic model is not only important for understanding the jet physics,
but also important for resolving the origin of the high-energy cosmic neutrinos, however, it is still very difficult for now.

\section*{Acknowledgments}
We thank the anonymous referee for insightful comments that have helped us improve the presentation
of the paper. DHY acknowledges funding support by China Postdoctoral Science Foundation under grant Nos. 2015M570152 and 2016T90136, by Key Laboratory of Astroparticle Physics of Yunnan Province (No. 2015DG035) and by the National Natural Science Foundation of China (NSFC) under grant No. 11573026. This work is partially supported by the National Natural Science Foundation of China (NSFC 11433004) and Top Talents Program of Yunnan Province, China. We acknowledge the financial support from the National Natural Science Foundation of China 11573060,
the Strategic Priority Research Program, the Emergence of Cosmological Structure of the Chinese Academy of Sciences, Grant No. XDB09000000.
SNZ acknowledges partial funding support by 973 Program of China under grant 2014CB845802, by the National Natural Science Foundation of China (NSFC) under grant Nos. 11133002 and 11373036, by the Qianren start-up grant 292012312D1117210, and by the Strategic Priority Research Program ``The Emergence of Cosmological Structures'' of the Chinese Academy of Sciences (CAS) under grant No. XDB09000000.

\bibliography{refernces}

\appendix
\section{Model parameters derived in the fits with the LP and BPL EEDs}

\begin{table*}
\begin{center}
 \caption{Model parameters derived in the one-zone SSC model with the LP EED. The mean values and the marginalized 68\% confidence intervals (CI) are reported. }
\label{inputlp}
\begin{tabular}{@{}cccccccccccc}
\hline
&  $\zeta_{\rm e}$  & $b$  & $B'$    & $t_{\rm var}$   &$\delta_{\rm D}$  &$\gamma'_{\rm pk}$\\
&                     &    &(0.1G)    &($10^4\ $s)                 &            & ($\rm 10^{4}$)\\
\hline
55265     &44.0       & 0.26    & 0.108                  & 8.00   &33.8        &2.88   \\
(68\% CI) &42.8-50.0  & 0.25-0.27 & 0.096-0.12      & -     &32.7-34.8   &2.58-3.16 \\
\hline
55266      &44.8         & 0.37     & 0.102                   & 8.00  &34.2   &4.91  \\
(68\% CI)  & 43.8-50.0  & 0.36-0.37  & 0.096-0.108      & -  &33.3-35.0   &4.68-5.15  \\
\hline
55267         & 19.4     & 0.56       & 0.252                   & 8.00  &27.2     &4.30  \\
(68\% CI)     & 0.01-50.0    & 0.55-0.57  & 0.072-0.396       & -   &21.3-33.2  &3.45-5.17 \\
\hline
55268         & 26.8    & 0.52       & 0.132                    & 3.00 &55.8      &4.09 \\
(68\% CI)    &19.17-50.0  & 0.51-0.53  & 0.06-0.18      & -     &47.5-63.9   &3.47-4.68 \\
\hline
55269         & 39.4       & 0.56       & 0.084               & 8.00  &39.8     &5.95 \\
(68\% CI)       &36.5-50.0    & 0.55-0.57  &0.072-0.096    & -   &37.6-41.9   &5.56-6.34\\
\hline
55270         & 30.1       & 0.71       &0.096                & 8.00   &39.4       &4.98 \\
(68\% CI)    &24.2-50.0  & 0.70-0.73 &0.06-0.12      & -   &35.0-43.6  &4.41-5.52 \\
\hline
55271         & 24.6     & 0.96       & 0.096                     & 8.00 &43.1     &7.62  \\
(68\% CI)    & 17-50.0  &0.92-1.00  &0.036-0.132      & -       &36.1-49.9 &6.43-8.82 \\
\hline
55272         & 3.6      & 0.83       & 0.816             & 8.00   &20.5      &3.85  \\
(68\% CI)   &0.01-2.9  & 0.81-0.84  &0.288-1.356    &  -        &14.3-26.5   &2.93-4.70\\
\hline
55273        &12.8           & 1.16     &0.108       & 8.00  &44.4   &7.42 \\
(68\% CI)    &0.01-14.1    & 1.12-1.20 &0.06-0.168     & -    &36.9-52.1 &6.12-8.74 \\
\hline
55274         &16.8        & 0.61       &0.42          & 8.00  &31  &2.67  \\
(68\% CI)       &0.01-50.0   & 0.59-0.62 &0.001-0.3    & -     &18.3-42.6  &1.67-3.59\\
\hline
55276         & 22.2        & 0.61       &0.12           & 8.00  &38.5    &3.64 \\
(68\% CI)  &0.01-50.0  & 0.60-0.62  &0.001-0.132         & -          &31.3-45.6  &2.96-4.30 \\
\hline
55277          &10.3  & 0.69       &0.18                  & 8.00     &35.2      &3.57     \\
(68\% CI)   &0.01-11.2 & 0.68-0.70   &0.084-0.252     & -    &28.9-41.7   &2.97-4.19  \\
\hline
\end{tabular}
\end{center}
\end{table*}

\begin{table*}
\begin{center}
 \caption{Model parameters derived in the one-zone SSC model with the BPL EED. The mean values and the marginalized 68\% confidence intervals (CI) are reported. }
\label{inputbpl}
\begin{tabular}{@{}ccccccccc}
\hline
MJD &    $\zeta_e$   & $p_2$   & $B'$       & $p_1$  & $t_{\rm var}$   & $\delta_{\rm D}$   & $\gamma'_{\rm c}$  \\
&                                 &           &(0.1 G)                      &       &   ($10^4\ $s)        &  &($\rm 10^{5}$) \\
\hline
55265         & 171        & 7.47            & 0.080            & 2.19     & 8.0     &34.6        &18.39  \\
 (68\% CI)       &165-200     &6.63-10    & 0.073-0.087   & 2.18-2.21  &-       &33.5-35.7    &17.53-19.22 \\
 \hline
55266         & 165          & 7.48       & 0.083              &  2.12         & 8.0       &33.9          & 12.98 \\
 (68\% CI)       & 157-200   & 6.62-10  & 0.010-0.084     &  2.11-2.13    &-          & 32.6-35.2    &12.42-13.53 \\
\hline
55267         & 84           & 7.77       & 0.165              &  2.13         & 8.0     &27.3          & 6.29 \\
 (68\% CI)       & 54-94   & 6.52-9.14  & 0.001-0.201     &  2.12-2.14     & -         & 16.0-28.7   &5.57-7.24 \\
 \hline
55268         & 117            &7.98       & 0.093               &  2.11         & 3.0     &55.0        &6.37 \\
 (68\% CI)       & 93-200    & 7.31-10  & 0.075-0.101     &  2.10-2.12     & -      &52.6-58.1   &5.00-6.74  \\
 \hline
55269         & 159           & 7.39       & 0.071               &  2.12         & 8.0    &37.9         &7.91 \\
 (68\% CI)       & 147-200    & 6.47-10  & 0.001-0.074     & 2.10-2.13     & -       &37.1-38.9     &7.47-8.31 \\
 \hline
55270         & 158          & 6.42        & 0.077             &  2.19       & 8.0    &37.1        &5.31  \\
 (68\% CI)       & 143-200    & 5.12-10  & 0.001-0.082  & 2.17-2.21   & -      &36.1-37.7   &5.03-5.59   \\
 \hline
55271         & 61            & 4.16       & 0.107              &  1.69         & 8.0    & 37.9          & 2.52  \\
 (68\% CI)       & 1-70    & 4.07-4.24   & 0.001-0.126     & 1.55-1.81     & -        & 32.9-40.8       &2.19-2.78  \\
 \hline
55272         & 67         & 7.45        & 0.110               &  1.97         & 8.0      &35.5           &4.20   \\
 (68\% CI)       & 36-72     &7.02-7.90    & 0.001-0.139    & 1.96-1.99     &  -     &30.5-38.5    &3.72-4.48 \\
 \hline
 55273        & 120            & 5.68      & 0.050             &  1.87         & 8.0    &45.5          &3.76  \\
 (68\% CI)       &102-200       &4.78-5.87  & 0.042-0.050     & 1.72-2.00     & -     &43.0-45.0     &3.17-4.25 \\
 \hline
55274         & 96           & 7.04       & 0.153               &  2.35         & 8.0    &30.5     &5.27 \\
 (68\% CI)       & 63-200     & 6.22-10  & 0.103-0.170     &  2.34-2.37     & -     &27.8-33.5  &4.84-5.79 \\
 \hline
55276         & 112            & 8.26       & 0.102              &  2.27        & 8.0      &36.0      & 5.76 \\
 (68\% CI)       & 73-200      &7.72-10  & 0.001-0.116     &  2.25-2.28       & -     &33.3-39.3  &5.33-6.31 \\
 \hline
55277          & 61           & 8.66      & 0.140             &  2.21        & 8.0     &33.5        &4.46 \\
 (68\% CI)       &28-73        & 8.32-9.11   & 0.001-0.190     & 2.20-2.22    & -     &28.7-36.4     &3.91-4.83 \\
 \hline
\end{tabular}
\end{center}
\end{table*}
\label{app}
\bsp	
\label{lastpage}
\end{document}